\documentstyle[multicol,prl,aps,graphics]{revtex}
\begin{document}
\title{Saturation of electrical resistivity in metals at large temperatures}
\author{M. Calandra and O. Gunnarsson}
\address{ Max-Planck-Institut f\"ur Festk\"orperforschung 
D-70506 Stuttgart, Germany}

\maketitle
\begin{abstract}
We present a microscopic model for systems showing resistivity
saturation. An essentially exact quantum Monte-Carlo calculation
demonstrates that the model describes saturation. We give a 
simple explanation for saturation, using charge conservation and
considering the limit where thermally excited phonons have destroyed 
the periodicity.  Crucial model features  are phonons coupling  
to the  hopping matrix elements and a unit cell with several atoms.            
We demonstrate the difference to a  model of alkali-doped 
C$_{60}$ with coupling to the level positions, for which there is 
no saturation. 
\end{abstract}
\begin{multicols}{2}
In a metal, the electrical resistivity, $\rho$, grows with 
the temperature $T$ due to the increased  scattering of the 
electrons by phonons. Typically, $\rho(T)\sim T$
for large $T$.   For some metals with a very large $\rho$, however, 
the resistivity saturates\cite{saturation,Fisk,Allen}, i.e.,
it grows very slowly with $T$ for large $T$. The resistivity 
is often described in a semiclassical (Boltzmann) picture,
where an electron, on the average, travels a mean free path $l$ before
it is scattered. The resistivity is inversely proportional to $l$.    
Typically, $l\gg d$, where $d$ is the atomic separation. For systems 
with resistivity saturation, however, $l$ becomes comparable to $d$.
Work in the 1970's suggested that resistivity saturation occurs 
universally when $l\sim d$, the Ioffe-Regel condition\cite{ioffe}, 
providing an upper limit to the large $T$ resistivity of metals.
Later work has, however, found exceptions, such as alkali-doped
fullerenes\cite{Hebard,Batlogg}. 

Intuitively, resistivity saturation seems natural. One might expect 
that at worst, an electron could be 
scattered at each atom, leading to $l\sim d$. Such a semiclassical 
picture, however, breaks down when $l\sim d$\cite{Kohn}, and it 
is contradicted by the lack of saturation for fullerenes. Several 
theories of the saturation have been presented, usually 
based on generalisations  of the semiclassical Boltzmann theory, but 
none has been generally accepted\cite{Chakraborty,Cote,Ron}.

The Bloch-Boltzmann theory starts from a periodic system and treats
the scattering mechanisms as small perturbations. Here we consider the
opposite limit, where thermally excited phonons have removed all 
effects of periodicity. In this limit, charge conservation naturally 
leads to saturation for systems with several atoms per unit cell
and strong electron-phonon coupling, in particular for systems 
where the phonons couple to the hopping matrix elements. 
We show that this happens to occur when $l \sim d$. This does not 
happen, however, for a model of alkali-doped C$_{60}$, where the 
phonons couple to the level positions. We first use an essentially 
exact quantum Monte-Carlo (QMC) calculation to demonstrate  
saturation in our model. We then introduce a method where the phonons 
are treated semiclassically and the electrons quantum-mechanically, 
justifying  the method by comparing with the QMC results. This method 
is sufficiently simple to allow for an interpretation of the results.     

Saturation is clearly observed for, e.g., 
A15 compounds, such as Nb$_3$Sb\cite{saturation}, while Nb metal 
shows weak saturation at large $T$\cite{Nb}.
We study a model Nb$_3^{\ast}$ of Nb$_3$Sb, where the 
Nb atoms have the same positions as in Nb$_3$Sb, but the Sb 
atoms are neglected\cite{Pickett}. This is compared with a model of Nb.
We consider a cluster of $N$ atoms, placed on  
A15 (Nb$_3^{\ast}$) or bcc (Nb) lattices with the     
lattice parameters 5.17 \AA \ (Nb$_3^{\ast}$) or  3.28 \AA \ (Nb).  

We study the scattering of the electrons from phonons. Each 
atom is assumed to have vibrations in the three coordinate 
directions, described by Einstein phonons. The phonon energy,
$\omega_{ph}=14$ meV, is obtained from an average over the phonons         
of  Nb\cite{Wolf}.  For each Nb atom we include the five-fold 
degenerate ($n=5$) $d$-orbital. The hopping matrix elements between 
the orbitals are obtained from Harrison\cite{Harrison}, using a 
power dependence on the atomic distance $d$ ($\sim 1/d^m$). We use 
$m=3.6$ more appropriate for Nb\cite{ove} than    
$m=5$ used by Harrison.  To avoid divergencies for very small $d$,
$1/d^m$ is replaced by $1/(d^m+a^m)$, with $a=2$ \AA. The atomic
vibrations modulate the hopping matrix elements, both due to the
changes of the atomic distances and the relative movements of the 
orbital lobes. We neglect the influence of the vibrations on the level
energies as well as the Coulomb interaction between the electrons. 
To obtain the 
resistivity, we calculate the current-current correlation function.
The current operator ${\bf j}$ is obtained by using charge and current 
conservation. Thus the matrix elements between orbitals $\nu$ and
$\mu$ on sites ${\bf R}^{\nu}$ and ${\bf R}^{\mu}$ are  
\begin{equation}\label{eq:0}
{\bf j}^{\nu\mu}=ie({\bf R}^{\nu}-{\bf R}^{\mu})t_{\nu\mu},
\end{equation}
where $t_{\nu\mu}$ are the corresponding hopping matrix elements.

This model can be solved essentially exactly by using a determinantal
quantum Monte-Carlo (QMC) method\cite{scalapino}, treating the 
phonons quantum mechanically. For the models 
studied here, the QMC method has no ``sign-problem''. The
calculated correlation function (for imaginary time) therefore 
has only (small) statistical errors. We use 
a maximum entropy method\cite{jarrell} to obtain the optical 
conductivity $\sigma(\omega)$ on the real frequency axis.
The resistivity $\rho$ is then $1/\sigma(0)$.

\begin{figure}
\centerline{
\rotatebox{0}{\resizebox{!}{2.3in}{\includegraphics{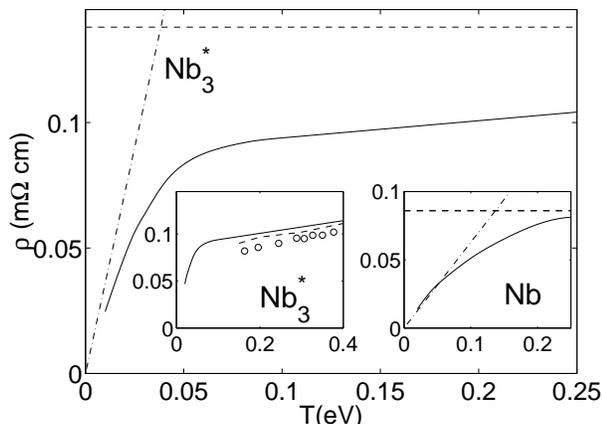}}}}
\caption[]{\label{fig:1}Resistivity $\rho(T)$ as  function of  
temperature $T$. The main figure shows results for Nb$_3^{\ast}$ 
($N=648$ atoms) and the right insert for Nb ($N=640$ atoms). 
In both cases the phonons are treated semiclassically. The small 
$T$ (Eq. (\ref{eq:4})) (chain curves) and the large $T$ (Eq. (\ref{eq:2a})) 
(broken curves) limits of $\rho(T)$ are also shown. The left insert 
compares the semiclassical (broken ($N=36$) and 
full ($N=648$) curves) and  quantum Monte-Carlo 
(circles, $N=36$) calculations for Nb$_3^{\ast}$. }
\end{figure}

The left insert of Fig. \ref{fig:1} shows  the QMC result (circles) 
for $\rho(T)$ of the Nb$_3^{\ast}$ model.      
We are particularly interested in the large $T$ behaviour, which is
also the limit where the QMC calculation can be performed with
a reasonable effort. The large $T$ result extrapolates to a
substantial nonzero value. However, since the resistivity of the 
Nb$_3^{\ast}$ model must go to zero for $T=0$, the QMC calculation 
clearly shows that the model leads to a drastic reduction of the slope 
of the $\rho(T)$ curve for large $T$, usually referred to as resistivity
saturation.

To analyse the results, we treat the phonons
(semi)classically. The atomic displacements due to the phonons are
chosen randomly according to a Gaussian distribution, determined         
from the average number 
of phonons at that temperature.  For given
``frozen'' displacements, we calculate the hopping and current 
matrix elements.  The eigenvalues $\varepsilon_i$ 
and eigenvectors $|i\rangle$ of the resulting     Hamiltonian are calculated. 
For an isotropic system, the optical conductivity is then given by
\begin{equation}\label{eq:1}
\sigma(\omega)\sim {1\over \omega}\sum_{ij}|\langle i|j_x|j\rangle|^2
(f_i-f_j)\delta(\hbar \omega-\varepsilon_j+\varepsilon_i),
\end{equation}
where $f_i$ is the Fermi function for the state $i$.                           

The left insert of Fig. \ref{fig:1} shows that this approach (broken line)
agrees quite well with the QMC calculation for large $T$, the temperature
range we are interested in, and we expect the semiclassical 
calculation to remain accurate for $T\gg\omega_{ph}$(=0.014 eV). We 
can therefore interpret the results by analysing the simpler 
semiclassical calculation. We observe that this model differs 
qualitatively from the (Ziman solution\cite{Grimvall}) of the 
Boltzmann equation, in which $\rho(T)\sim T$ for large $T$.
In the Boltzmann equation the electrons are treated semiclassically,
while here they are treated quantum-mechanically.                

Fig. \ref{fig:1} shows that there is a very clear saturation for 
Nb$_3^{\ast}$, while for Nb (right insert) there is only weak 
saturation at large $T$, in agreement with experiment\cite{saturation,Nb}.

\begin{figure}[bt]
\centerline{\rotatebox{0}{\resizebox{!}{2.15in}{\includegraphics
{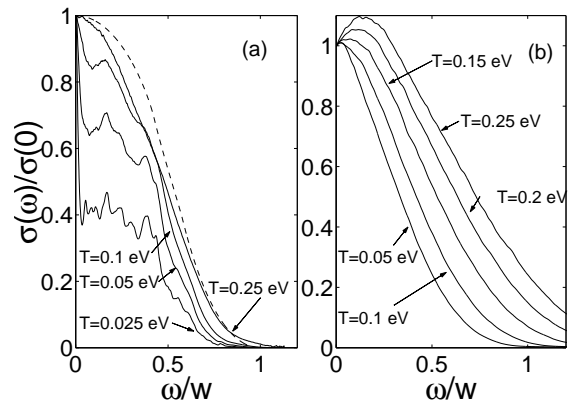}}}}
\caption[]{\label{fig:2}Optical conductivity as  function of  
frequency $\omega$ for the 
(a) A15 and (b) fullerene models in the semiclassical calculation. 
The frequency has been scaled  by the $T=0$ band width $W$.
(a) also shows (broken curve) the result of approximating all current
matrix elements by their average.  }
\end{figure}

We now discuss this saturation. Fig. \ref{fig:2}a shows 
$\sigma(\omega)$ for Nb$_3^{\ast}$ in        
 the semiclassical theory. For small $T$, there is a narrow Drude peak 
at $\omega=0$, which is smeared out for 
large $T$. Then $\sigma(\omega=0)$ drops 
correspondingly. For $\omega>W$, $\sigma(\omega)$ remains zero 
(apart from a slight broadening introduced in the
calculation),  where $W$ is the band width, since there are no 
excitations for $\omega>W$. 

\begin{figure}[bt]
\centerline{\rotatebox{-90}{\resizebox{!}{3.00in}{\includegraphics
{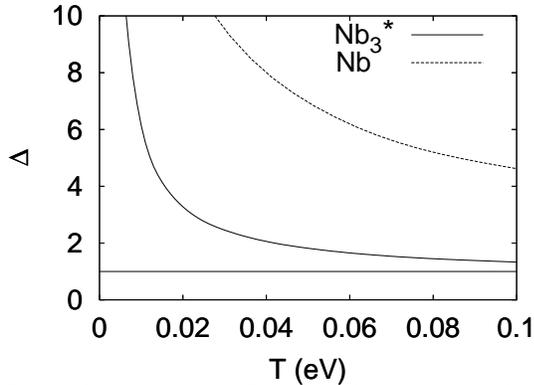}}}}
\caption[]{\label{fig:3}The measure $\Delta$ (Eq. (\ref{eq:2d})) of 
the amount of momentum conservation as a function of $T$ for Nb$_3^{\ast}$
(full line) and Nb (broken line). 
The horizontal line represents complete loss of momentum 
conservation. 
 }
\end{figure}

The Drude peak is due to intraband transitions between states 
with similar values of the wave vector ${\bf k}$. As $T$
is raised and the vibrations of the atoms are increased,
the states loose their well-defined ${\bf k}$ and band index  
labels. To illustrate this, we  decompose a state $|i\rangle$ at 
a finite $T$ in  the $T=0$ states with given ${\bf k}$-vectors.
The corresponding weights are labelled $c(i{\bf k})$. We define 
$\Delta$ as the average of $\Delta(i)$ over all states $i$, where
\begin{equation}\label{eq:2d}
\Delta(i)=n_{\bf k}\sum_{\bf k}c(i{\bf k})^2,
\end{equation}
and  $n_{\bf k}=256$ is the number of allowed ${\bf k}$-vectors 
for the systems studied.  If each state has $n_{\bf k}/m$ 
${\bf k}$-components
with equal weights, $\Delta=m$. If periodicity is completely
lost, $\Delta=1$, and if each state has only one ${\bf k}$-component,
$\Delta=n_{\bf k}(=256)$\cite{T=0}.  
$\Delta$      is shown in Fig. \ref{fig:3}. It illustrates how
the effects of periodicity are lost very quickly for Nb$_3^{\ast}$
on a temperature scale of just a few hundred K.                         

For a large $T$, the (${\bf q} \to {\bf 0}$) current operator then 
couples all states to each others. We now make the assumption 
that at large $T$ the coupling between all states is equally 
strong\cite{current}. This is the opposite limit to the Boltzmann 
treatment, where ${\bf k}$ is assumed to be a good quantum 
number. We replace a current matrix element in Eq. (\ref{eq:1}) 
by its average $| j_x^{\rm av}| ^2$ over all transitions.

\noindent
Using charge 
and current conservation (Eq. (\ref{eq:0})), we obtain
\begin{equation}\label{eq:2e}
| j_x^{\rm av}| ^2 =  
{e^2d^2\over 3N^2n^2} \sum_{\nu\mu} |t_{\nu\mu}|^2
={e^2d^2\over 3Nn^2} \langle \varepsilon^2\rangle,
\end{equation}
where $n$ is the orbital degeneracy, $\langle \varepsilon^2\rangle$ 
is the second moment of the density of states $N(\varepsilon)$ per atom
and $d$ is a typical atomic separation.
We assume some generic shape of $N(\varepsilon)$. Since we know its 
second moment, we can then relate $N(\varepsilon)$ to the $j_x^{\rm av}$. 
Using this  relation, we can calculate $\sigma(\omega)$,               
shown by the broken curve in Fig. \ref{fig:2}a. A comparision  
with the full calculation shows that this approximation becomes
rather accurate already at a moderate $T$. We  find that 
$\sigma(0)\sim d^2/\Omega$, where 
$\Omega$ is the volume per atom. Using the $T=0$ nearest neighbour 
distance for $d$, we obtain 
\begin{equation}\label{eq:2a}
\rho \sim A {d\over n}.
\end{equation}
$A$ depends on the precise assumptions of the model, but it is $\sim0.2$  
m$\Omega$cm if $d$ is expressed in \AA. Assuming a semi-elliptical 
density of states $\sim \sqrt{(W/2)^2-\varepsilon^2}$ and band filling 
0.4, we find $A=0.27$ and $0.15$ m$\Omega$cm for the A15 and bcc lattices, 
respectively.  This is shown by the broken horizontal lines in Fig. 
\ref{fig:1}. Given the simple assumptions, the agreement with the full 
semiclassical calculations at large $T$ is surprisingly good.
The corresponding apparent mean-free path is 
\begin{equation}\label{eq:3}
l\sim c n^{1/3} d.
\end{equation}
With the assumptions above, we obtain $c=0.5$ and $c=0.6$ for the 
A15 and bcc lattices, respectively. Saturation therefore happens roughly when
the Ioffe-Regel condition is satisfied, as might have been expected
on dimensional grounds. 

The derivation of Eq. (\ref{eq:3}) uses a quantum-mechanical treatment
of the electrons.  It explains why metals with a large 
resistivity usually show saturation, and why it   happens when 
$l \sim d$.

The $\rho(T)$ in Eq. (\ref{eq:2a}) is independent of $T$,
while Fig. 1 shows a weak $T$ dependence even after
saturation. This is partly due to the $T$-dependence of the
Fermi-functions in Eq. (\ref{eq:1}), which were approximated  
by $\theta$-functions in the derivation above.  There is 
also a $T$ dependence due to  changes of the {\it shape} of 
the density of states as $T$ is increased, which was neglected above  
by using a ``generic'' density of states. These effects
are only important when $T$ is an appreciable fraction of 
the band width (for realistic values of the electron-phonon
coupling). 

Eqs. (\ref{eq:2a}, \ref{eq:3}) are obtained by assuming that 
the Drude peak has been smeared out and that $\sigma(\omega)$
is spread out over the whole band width. Both the saturation
resistivity and the corresponding mean free path are
independent of the band width.
A scaling of $t_{\mu\nu}$ by a factor $\alpha>1$ increases $W$,
and $\sigma(\omega)$ extends over a larger energy range.
At the same time, however, $j_x^{\rm av}$ is increased (Eq. 
(\ref{eq:0})) in such a way that $\sigma(0)$ in Eq. (\ref{eq:1})
is unchanged. Charge and current conservation therefore plays a 
crucial role for our results (\ref{eq:2a},\ref{eq:3}) .

We next consider a small $T(>\omega_{ph})$. Then\cite{Grimvall}
\begin{equation}\label{eq:4}
\rho(T)= 8\pi^2{\lambda T k_B\over \hbar \Omega_{pl}^2},
\end{equation}
where $\lambda$ is the electron-phonon coupling constant and
$k_B$ is the Boltzmann constant. $\Omega_{pl}$ is the plasma frequency 
\begin{equation}\label{eq:5}
(\hbar\Omega_{pl})^2={e^2\over 3\pi^2}\sum_n \int_{Bz}d^3k \lbrack {\partial
\varepsilon_{n{\bf k}} \over \partial {\bf k}}\rbrack^2\delta
(\varepsilon_{n{\bf k}}-E_F).
\end{equation}
where $\varepsilon_{n{\bf k}}$ is the energy of a state with
the band index $n$ and the wave vector ${\bf k}$ and $E_F$ is
the Fermi energy. $\Omega_{pl}$ depends on the average Fermi velocity.                

The straight line given by Eq. (\ref{eq:4}) (chain lines in  Fig. 
\ref{fig:1}), agrees well with the semiclassical calculations 
for small $T$. Typically, this line rises so slowly that it 
intersects the horizontal line of Eq. (\ref{eq:2a}) well above 
the melting point for the metal of interest. Then no saturation
is found in the accessible temperature range. If, however, 
$\Omega_{pl}$ is small and $\lambda$ is large, the two lines 
cross below the melting temperature. Saturation is then  
observed. An important difference between Nb$_3^{\ast}$ and Nb
is that  Nb$_3^{\ast}$ has many rather flat bands, due to the 
large unit cell\cite{Chakraborty} and many forbidden crossings. 
The resulting small
electron velocities, lead to a small $\Omega_{pl}$. We find 
$\Omega_{pl}= 3.6$ and 8.2 eV for Nb$_3^{\ast}$ and Nb,
respectively, which makes the slope of the line in Eq. 
(\ref{eq:4}) about a factor of five larger for Nb$_3^{\ast}$
than for Nb. We find similar $\lambda$'s for 
Nb$_3^{\ast}$ ($\lambda=1.0$) and  Nb ($\lambda=0.9$).              
More accurate estimates give $\Omega_{pl}=3.4$ eV 
(Nb$_3$Sn)\cite{Mattheiss} and 9.5 eV (Nb)\cite{Savrasov}
and $\lambda=1.7$ (Nb$_3$Sn)\cite{Allen} and 1.1 (Nb)\cite{Savrasov}.

It is interesting to replace the $d$-orbitals in our Nb$_3^{\ast}$
model by $s$-orbitals. As before, the resistivity shows saturation,
but the saturation is a less pronounced than in Fig. 1. The 
saturation resistivity  (Eq. (\ref{eq:2a})) is larger due to the 
smaller degeneracy ($n=1$). 

A very different behaviour is found in the alkali-doped
fullerenes, (A$_3$C$_{60}$ (A= K, Rb)), where the apparent    
mean free path becomes much shorter than the separations of the 
C$_{60}$ molecules\cite{Hebard,Batlogg}. This behaviour was related 
to the fact that the important (intramolecular) phonons primarily 
couple to the C$_{60}$ level energies, instead of the hopping matrix 
elements\cite{Nature,sat}. Already at a moderate $T$,          
the resulting fluctuations in the level energies become comparable to 
the $T=0$ width of the narrow $t_{1u}$ band, which conducts the current. 
This leads to 
a   broadening of the $t_{1u}$ band and of $\sigma(\omega)$ beyond 
the $T=0$ band width. In contrast to the case of a scaling of the 
hopping parameters, discussed below Eq. (\ref{eq:3}), this is, however,
not accompanied by an increase of the current operator. Thus $\sigma(0)$ 
is reduced, explaining the lack of saturation. This is illustrated in 
Fig. \ref{fig:2}b. 

It is interesting to compare the cases when the phonons couple
to the level energies (LE coupling) and to the hopping integrals
(HI coupling).  We have studied the case of HI coupling in a 
C$_{60}$ model. By assuming an
unrealistically small phonon frequency for the intermolecular 
vibrations, we can obtain the same value of $\lambda$ as for 
the LE coupling. For the HI coupling, we then find  that the 
resistivity our C$_{60}$ model shows saturation. 
We have also considered LE coupling in the Nb$_3^{\ast}$ 
model. The resulting resistivity  shows a change in slope, 
somewhat similar to Fig. 1, but with a larger 
slope for large $T$ than in Fig. 1.  
These results illustrates that it is possible to obtain 
saturation with LE coupling, but that HI coupling is
more appropriate for describing saturation.  

Our semiclassical treatment of the two models is closely related to
conduction in disordered systems with diagonal (DD) or  (ODD) 
off-diagonal disorder. While DD models can give localization,
no localization was not found close to the center of the band in a ODD 
model\cite{offdiagonal}.  This is consistent with the saturation 
seen in Fig. 1 for the A15 model, having ODD in the semiclassical
treatment. For the C$_{60}$ model, the semiclassical treatment gives 
localization for a large DD. The QMC treatment, however, 
takes  into account that the 
disorder is not static but due to thermal fluctuations and that
the scattering can be inelastic. In this QMC treatment we find a lack
of saturation, but we have not seen signs of localization.

To summarize, guided by the loss of periodicity at large $T$, we 
have studied the effect of replacing the current matrix element 
by its average.  Together with charge conservation, this leads 
to clear saturation in a model where the phonons couple to the hopping 
matrix elements. On the other hand, saturation does not happen in 
a model for A$_3$C$_{60}$, where the phonons couple to the level 
energies. The issue of saturation or not saturation is only raised 
for experimentally accessible temperatures if $\lambda$ is 
large and $\Omega_{pl}$ is small. This is 
favoured by the relatively flat bands for Nb$_3^{\ast}$ ($\Omega_{pl}$ 
small) and by the small band width for A$_3$C$_{60}$ ($\lambda$ 
large and $\Omega_{pl}$ small).

We would like to thank M. F\"ahnle, O. Jepsen, E. Koch and R. Zeyher for 
useful discussions, M. Jarrell for making his MaxEnt program available
and the Max-Planck-Forschungspreis for financial support.

\end{multicols}

\begin{thebibliography}{*}
\bibitem{current}
The matrix elements of the current operator have a broad
distribution of values. However, an average over states with the 
similar energies, appropriate in Eq. (\ref{eq:1}), has fairly small 
variations already at moderate $T$.
\bibitem{sat}        
A similar model was used to describe resistivity saturation\cite{Millis}. 
We find,  however, that this model is more appropriate for describing 
lack of saturation.
\end{thebibliography}
\end{document}